**Inferring Global Dynamics Using a Learning Machine**

Hong Zhao
Department of Physics, Xiamen University,
Xiamen 361005, China
E-mail: zhaoh@xmu.edu.cn

**Given a segment of time series of a system at a particular set of parameter values, can one infers the global behavior of the system in its parameter space? Here we show that by using a learning machine we can achieve such a goal to a certain extent. It is found that following an appropriate training strategy that monotonously decreases the cost function, the learning machine in different training stage can mimic the system at different parameter set. Consequently, the global dynamical properties of the system is subsequently revealed, usually in the simple-to-complex order. The underlying mechanism is attributed to the training strategy, which causes the learning machine to collapse to a qualitatively equivalent system of the system behind the time series. Thus, the learning machine opens up a novel way to probe the global dynamical properties of a black-box system without artificially establish the equations of motion. The given illustrating examples include a representative model of low-dimensional nonlinear dynamical systems and a spatiotemporal model of reaction-diffusion systems.**

## I. Introduction

Dynamical systems are the basic objects of scientific research. They have broad applications in a variety of disciplines ranging from physics, biology, chemistry, engineering, economics, to medicine and beyond. The typical approach for understanding a dynamical system is to establish its equation of motion, which can expressed as $\boldsymbol{x}(t) = f^t(\boldsymbol{x}(0), \Lambda)$, where $\boldsymbol{x} \in R^d$ is a $d$-dimensional vector variable, $t$ is the time, $f^t$ is the evolution operator, and $\Lambda$ is the parameter set. By investigating the entire parameter space one can grasp the global properties of the system. The inverse-problem approach, one of the most important mathematical tools in science, aims to fix the equation of motion by finding its parameters from a data set of observations [1-6]. However, to accurately model a real-world system is an extremely difficult problem – it needs accurate prior knowledge about the system.

Without knowing the details of the model, for a particular variable of the system, $x(t)$, if it is measured over a period of time, there are some model-free approaches that can be applied to predict the future evolution of this variable, such as the time series reconstruction method [7-8], empiricism-based method [9], deep belief network [10], long short-term memory network [11], and reservoir computing [12-14], etc. This kind of prediction is done, however, for variable states with a *fixed* system parameter set $\Lambda$. A more challenging task is to probe the *global* dynamical behavior over a wide parameter space of the system *behind* the time series in a model-free way under the same premise, i.e., to infer the dynamical properties that the system may exhibit in its parameter space, by using only time series measured at a particular parameter set $\Lambda$.

Here we report that by employing a learning machine, this task can be fulfilled to a certain extent. We find that for an appropriate training strategy, the dynamics of the learning machine observed at different training stages may qualitatively reproduce the global dynamics of the system behind the time series (we call the target system hereafter) along special paths in its parameter space around the parameter set where the training time series is collected. These paths are particularly important since the primary dynamical behavior of the target system may be qualitatively reproduced along that, usually in a way from simple to complex.

The mechanism is also investigated. We show that our particular training strategy could collapse the parameter space of the learning machine rapidly onto a subspace in which the learning machine appears to be qualitatively equivalent to the target system. The parameter region reached in this early stage usually corresponds to the simplest dynamics. With further training, the learning machine evolves to the target system with the given parameters, and may extend to neighboring parameter regions, where more complex dynamical behaviors could be detected. In this way, the learning machine reveals the dynamics of the target system from simple to complex.

The paper is organized as follows. Section II introduces the learning machine we designed and shows how to train it. We emphasize the application of a gradient-free training method, i.e., a Monte Carlo algorithm. The advantages of this algorithm is that it enables us to adjust all parameters of the learning machine, using various neuron transfer functions, and guarantee the cost function decreases monotonically. Section III validates our strategy with the paradigmatic Lorenz system. Using this model, we illustrate to what an extent the global dynamics of the target system can be revealed. Section IV is contributed to show that our method is applicable also for spatiotemporal systems. The so-called Barrio-Varea-Aragon-Maini (BVAM) model [15-16] is employed, which represents a reaction-diffusion system with two species producing temporal oscillations and spatiotemporal chaos. This system exhibits rich dynamical behaviors arising from the interaction between Turing pattern and Hopf bifurcations [17,18]. Besides, it develops to chaos by the Ruelle-Takens-Newhouse route [19], instead of the usual period-doubling route as the Lorenz system follows. Section V explains the mechanism why the global dynamics of the target system can be reproduced by our learning machine. The last section summarizes the main results.

## II. Learning machine and training strategy

The dynamics of our self-evolution learning machine can be constructed by various dynamical equations. In this paper two of them are considered. One is a three-layer time-delayed map

$$\phi(n+1) = \phi(n) + \tau \sum_{i=1}^{N} u_i f\left(\beta_i \left(\sum_{j=1}^{M-1} v_{ij} \phi(n-j) - b_i\right)\right), \qquad (1)$$

where $M$ and $N$ are the number of input and hidden layer neurons, respectively, $\tau$ is the time interval between two sequential records of the target system's time series, and $f$ is the neuron transfer function. Besides $M$ and $N$, this learning machine contains four types of parameters, each is bounded in an interval, i.e., $|u_i| \leq c_u, |\beta_i| \leq c_\beta, |v_{ij}| \leq c_v,$ $|b_i| \leq c_b$. We call $c_u, c_\beta, c_v, c_b$ control parameters. The parameters of the learning

machine are randomly initialized in their value ranges. The role of the control parameters is to control the input-output sensitivity of the learning machine, similar to that of regularizations in conversional training algorithms. Let $x(j), j = 1, \ldots, P$ be a segment of time series of variable *x* of the target system at its given parameter set. We use it to construct *P-M* training samples, with $\{\phi(n-j) = x(n-j), j = M-1, \ldots, 0\}$ being the input of the *n*th sample and $x(n+1)$ the expected output. We define the cost function as

$$\lambda = \frac{1}{P-M}\sum_{n=M}^{P-M}(x(n+1) - \phi(n+1))^2. \tag{2}$$

This is the most important parameter in our framework. The global dynamics of the target system behind the training data emerges with the decrease of this parameter. Note that this learning machine is for reproducing the dynamics when only an one-dimensional time series of the target system is available.

Another learning machine is a three-layer map

$$\boldsymbol{\phi}(n+1) = \boldsymbol{\phi}(n) + \tau \sum_{i=1}^{N} u_i f\left(\beta_i(\hat{\boldsymbol{v}} \cdot \boldsymbol{\phi}(n) - b_i)\right), \tag{3}$$

where $\boldsymbol{\phi}(n)$ is an *M*-dimensional vector. The training samples are constructed by {input: $\boldsymbol{\phi}(n) = \boldsymbol{x}(n)$, output: $\boldsymbol{x}(n+1)$}, and the cost function is defined by

$$\lambda = \frac{1}{P}\sum_{n=1}^{P}\sum_{j=1}^{M}\left(x_j(n+1) - \phi_j(n+1)\right)^2. \tag{4}$$

This learning machine applies for the situation when all variables of the target system are available. In real applications, obtaining all of the system variables is usually a difficult task. Therefore, one can expect that the learning machine (1) should be more useful. In this paper, we employ (3) only for the sake of simplicity to explain the mechanism why a learning machine can infer the global dynamics of the target system. Note that *t=nτ* gives the time of the target system.

To perform the training, we choose a proper set of the control parameters $c_u, c_\beta, c_v, c_b$ for fixed *M* and *N*, and randomly initialize all of the parameters of the learning machine accordingly (i.e., they are bounded in the intervals of the control parameters). Then, randomly mutating a parameter in its value range and accepting this variation if it does not increase *λ*. Each adaptation renews only *O(P)* multiply-add operations and the adaptation accepted is optimum for the whole training set in the statistical sense. It does not need to evolve the entire network which needs about *O(NMP+NP)* multiply-add operations [20,21]. As such, this algorithm is practical for usual applications. This is different from the conversional gradient-descent algorithms. We call it the Monte Carlo algorithm, one advantage it has that is essential for our purpose is that it can decrease *λ monotonically*, making *λ* be the indicating parameter of the learning machine and along which the global dynamics of the target system emerges. Another advantage of the Monte Carlo algorithm is that one can employ a variety of neuron transfer functions.

In principle, there are no restrictions on the choice of the neuron transfer function. In this paper, unless state explicitly, we adopt $f(h) = \exp(-h^2)$ to be the neuron transfer function. The parameter *N* controls the size of the learning machine. It should be sufficiently large so that the results are insensitive to its particular value. Throughout this paper we take $N=3000$ for all of our examples.

## III. Examples
### III.1 The Lorenz system

The Lorenz system is given by $dx/dt = -\sigma(x - y)$, $dy/dt = -xz + Rx - y$, and $dz/dt = xy - Bz$, where $\sigma$, $R$, and $B$ are system parameters. In Fig. 1(a) we show a segment of time series of $x$ at $(\sigma, R, B) = (10, 28, 0.555)$. It is a period-2 limit cycle. We apple a time series recorded with the interval of $\tau=0.01$ over a time period of $t=30$ to train the learning machine (1). The control parameters are $c_\beta = 0.5$, $c_u = 0.2$, $c_v = 2$, $c_b = 15$. The time delay parameter is set to be $M=60$.

The global dynamical behavior of a system can be represented by the bifurcation diagram. We show the bifurcation diagram of the learning machine as follows: After a fixed amount of Monte Carlo operations, the training is suspended and the value of $\lambda$ is calculated following Eq. (2). Then, we input the last 60 data points ($\tau M=0.6$) of the record of $x(t)$ to evolve the learning machine. After the learning machine has been self-evolved for a time long enough to go beyond the transient process, we record a sufficient long time series of $\phi(n)$. This series represents the asymptotic dynamical behavior of the learning machine at the obtained $\lambda$ value. The delayed coordinates, $\phi(t_c - 0.01)$, of the section point $t_c$ that the $\phi$ variable crosses the line of $\phi=5$ from the above are then obtained ($t_c$ is obtained by nonlinear fitting). Fig. 1(b) shows the bifurcation diagram obtained in such a way as the function of $1/\lambda$. It can be see that the learning machine shows a complex dynamical behavior with the decrease of $\lambda$; it develops to chaos through the period-doubling bifurcation route.

The question of higher interest is whether the bifurcation diagram of the learning machine reveals knowledge of the target system. To answer this question, we plot the bifurcation diagram of the Lorenz system versus the parameter $B$ with other two parameters being fixed at $(\sigma, R) = (10, 28)$. The vertical axis represents $x(t_c-0.01)$, which is the delayed coordinate of the section point $t_c$ when the $x$ variable crosses the line of $x=5$ from the above. We see that the bifurcation diagram of the learning machine is highly similar to that of the Lorenz system in the interval of $B \in (0.45, B_c)$. To verify that they do intrinsically similar with each other, we make a comparison of trajectories at different stages of the two bifurcation diagrams. Figure 1(d) shows the time series $\phi(t)$ of the learning machine and $x(t)$ of the Lorenz system at several positions marked by digits along the bifurcation diagrams. Both the Lorenz system and the Learning machine are started up by in putting a segment of $x$ variable with length$\tau M=0.6$. We see that the pairs at corresponding stages have very high similarity. Meanwhile, they are quantitatively different except the pair marked by the digit 2. Before the point marked by the digit 2, the amplitude of $\varphi(t)$ is large than $x(t)$, and afterwards the amplitude of $\phi(t)$ is smaller than $x(t)$. Note that the digit 2 in Fig.1 (c) marks the point where the training sample is made. Therefore, the learning machine can quantitatively mimic the target system at the parameter set that the training sample is made, and can reproduce qualitatively the dynamical properties at nearby region in the parameter space. Though qualitatively, it provides us the essential knowledge of the target system. In our present example, for instance, the only information of the target system we know is a segment of its periodic trajectory shown in Fig. 1(a). The learning machine tells us that the target system behind the periodic trajectory may show complex behavior, i.e., it can evolve from simple limit cycle to chaotic motion through period-doubling bifurcation if one changes its system parameters properly. We call this function of the learning machine as the system evolution prediction.

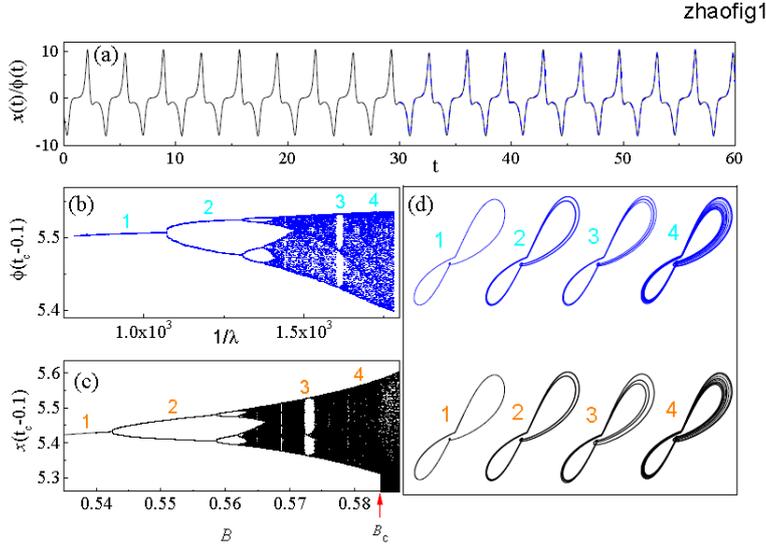

**Fig. 1.** (a) The time series of *x* variable (solid line) at *B*=0.555, and the predicted time series of $\phi(t)$ (dashed line). (b) The bifurcation diagram of the learning machine. (c) A segment of the bifurcation diagram of the Lorenz system. (c) Typical trajectories along the two bifurcation diagrams. They are represented by $x(t) - x(t - 0.1)$ and $\phi(t) - \phi(t - 0.1)$ with scales of $x(t)/\phi(t) \in (-15,15)$, respectively. The digits tag the points in the parameter space of these trajectories.

Figure 1(b) indicates that based only on a segment of *x* variable collected at the parameter point *B*=0.555, the learning machine can infer that the target system may evolve to chaos via the period-doubling route. However, the bifurcation diagram ends at the dynamical phase that corresponds to the crisis point of *B*=$B_c$. Further decrease the cost function induces a quick divergence or a sudden change in the amplitude of the output, i.e., the output of the learning machine no longer maintains a clear similarity with that of the target system. We have checked that applying data samples from any position of $B \in (0.45, B_c)$ would result in qualitatively similar bifurcation diagrams. Fig. 2 (a) and 2(b) show the bifurcation diagrams of learning machines trained by the time series of *x* variable measured at *B*=0.5725 and *B*=0.58, respectively. At *B*=0.5725 the Lorenz system shows a period-3 limit circle embedded in the chaotic region. The delayed-coordinate trajectory is shown in Fig. 1(d) and marked by digit 3. At *B*=0.58 the Lorenz system shows the chaotic motion, whose delayed-coordinate trajectory is marked by digit 4 in Fig. 1(d). We see that in both cases the global dynamics of the target system in the region of *B*<$B_c$ is reproduced, and the bifurcation diagram is qualitatively identical with that obtained by the training set of *B*=0.555. We have checked that no matter how close the sampling position is to $B_c$, the resulted bifurcation diagram is qualitatively the same and only the part similar to that of the Lorenz system with $B<B_c$ is shown up. Note that at the crisis point of $B=B_c$ the two coexisting-attractor branches collide (see Fig. 4) and thus the amplitude of the trajectories shows a sudden burst. Therefore, the reproducing ability of the learning machine is limited by intrinsic properties of the target system, as the crisis does that can abruptly increase the complexity. This property of the learning machine provides us a possible way to infer the crisis of the target system.

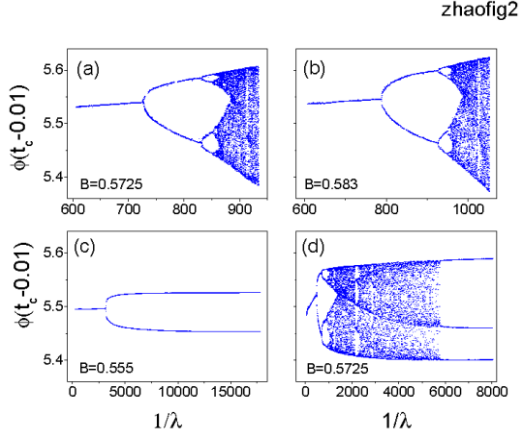

**Fig. 2.** Bifurcation diagrams of learning machines for training samples made at $B=0.5725$ (a), $B=0.583$ (b), $B=0.555$ (c), and $B=0.5725$ (d), respectively. Parameters used in (a) and (b) are $c_\beta = 0.45$ and $c_\beta = 0.4$ respectively and others are fixed at $c_u = 0.2, c_v = 2, c_b = 15$. Parameters used in (c) and (d) are $c_\beta = 0.4$ and $c_\beta = 0.3$ respectively and others are fixed at $c_u = 1, c_v = 1, c_b = 20$.

The global dynamics can be revealed depends also on the choice of control parameters. There are various combinations of training control parameters that can lead to the fully developed bifurcation diagram as Fig. 1(b), Fig. 2(a) and Fig. 2(b) show. Meanwhile, many other combinations may only partially reproduce the bifurcation diagram, of which the commonly encountered scenario is that the part from the period-one limit cycle to the present state that the training data is collected, as Fig. 2 (c) and 2(d) show. In these situations, qualitatively, the dynamics having lower level of complexity than that at the parameter set where the training data is collected is fully explored. In Fig. 2(c), the training data is the period-2 limit cycle at $B=0.555$, and only the evolution history from the period-1 limit cycle to this period-2 solution by a period-doubling bifurcation is explored. In Fig. 2(d), the training data is the period-3 limit cycle at $B=0.5725$, and the entire evolution history of the bifurcation is revealed till to the state of the training data.

To explore the dynamics in a wider region in the parameter space of the Lorenz system we need to apply the data collected at the parameter region with high-order complexity. In Fig. 3 (a) we show a segment of time series of $x$ at $(\sigma, R, B) = (10,28,2)$. Fig. 3(b) shows the bifurcation diagram of the learning machine resulted by this time series. The control parameters are $c_\beta = 0.2, c_u = 0.2, c_v = 2$, and $c_b = 15$. Fig. 3(c) zooms in the beginning part of the bifurcation diagram. In Fig. 3(d) we show also the bifurcation diagram of the Lorenz system in the region of $B \in (0.35, 2.2)$ with other two parameters fixed at $(\sigma, R) = (10, 28)$. We see that the global dynamics of the Lorenz system throughout the parameter space is very complicated. There are many stages with coexisting attractor branches. Several pairs of coexisting attractor branches are shown in the bifurcation diagram. Figure 1(c) is indeed an enlargement of the upper branch in the interval of $B\in(0.45, 0.587)$ in Fig. 3(d).

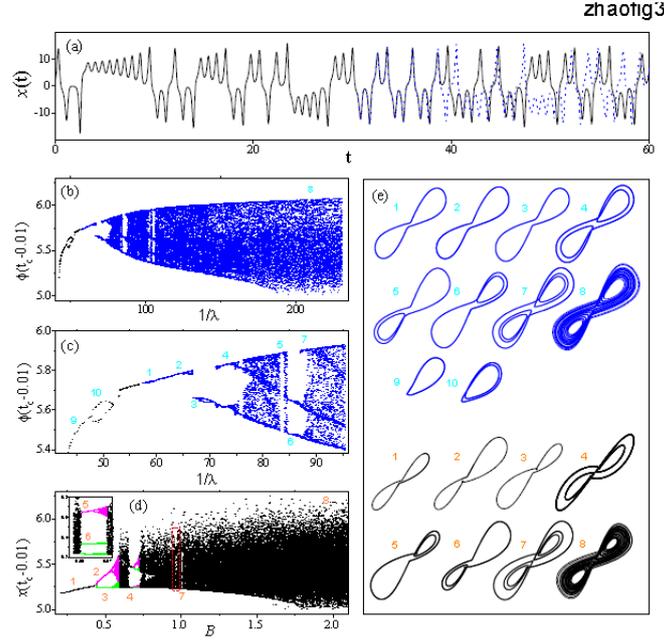

**Fig. 3.** (a) The time series of $x$ variable (solid line) at $B=2$, and the predicted time series of $\phi(t)$ (dashed line). (b) The bifurcation diagram of the learning machine, and (c) zooms in the beginning part. (d) The bifurcation diagram of the Lorenz system. The pieces colored in pink and green are coexisting attractor branches at the same parameter region. The inset is a zoom-in of the area marked by the red rectangle. (e) Typical trajectories along the two bifurcation diagrams. They are represented by $x(t) - x(t - 0.1)$ and $\phi(t) - \phi(t - 0.1)$ with scales of $x(t)/\phi(t) \in$ (-20,20), respectively. The digits tag the positions of these trajectories.

It seems that Fig. 3(b) and Fig. 3(c) have obvious differences with Fig. 3(d). However, on one hand, the learning machine at the optimal $\lambda$ can also mimic the dynamics of the target system at the parameter set where the training sample is made. The trajectories in Fig. 3(e) marked by the digit 8 show the sample time series of $x(t)$, and the time series of $\phi(t)$ of the learning machine at $1/\lambda=220$, both are started up by the same input segment of $x(t)$ of length $\tau M=0.6$. We see that they have high similarity. The predicted time series of $\phi(t)$ is also shown in Fig. 3(a). It indicates that the learning machine can trace the target system up to t=10. Hereafter, the two trajectories separate with each other but still keep the similar feature of motion. Furthermore, by examining the trajectories we can confirm that those around $B=2$ of the bifurcation diagram of the Lorenz system, and those around digit 8 in the bifurcation diagram of the learning machine have high similarity.

On the other hand, even far from the region of digit 8, we find that the periodic solutions along the bifurcation diagram of the learning machine reproduce the primary limit cycles appeared in period windows along the bifurcation diagram of the Lorenz system, see Fig. 3(e). These solutions are distributed in the region of $B<1$, far from $B =$

2 at which the training data is collected. An extra segment in the beginning of the bifurcation diagram of the learning machine shows up and two trajectories, marked by digit 9 and 10 on this segment, are also shown in Fig. 4(e). Interestingly, by carefully checking the Lorenz system, we do find that there are solutions of the Lorenz system around $B=0.05$.

Meanwhile, the learning machine loses certain detailed information of the target system. The primary limit cycles remain but, taking the two period-1 limit cycles, marked by 2 and 7, for example, the subsequent period-doubling cascade that has appeared in the bifurcation diagram of the target system is lost. In addition, the coexisting-attractor branches that appear simultaneously in the bifurcation diagram of the target system may appear discontinuously along the axis of $1/\lambda$ (see Fig. 3(c)), implying that the symmetry of the target system is not precisely kept. In addition, from Fig. 3(e), we see that the amplitudes of the predicted trajectories are inconsistent with those of the corresponding trajectories of the Lorenz system, except that for the current parameter set. The reproduction of the global dynamics is thus only qualitative.

## III.2 The BVAM model
The BVAM model is given by [15]
$$\frac{\partial u}{\partial t} = 0.08\nabla^2 u + (u - v - Cuv - uv^2),$$
$$\frac{\partial v}{\partial t} = \nabla^2 v + \left(-\frac{3}{2}v + 3u + Cuv + uv^2\right).$$

Here we keep only the parameter $C$ free while others the same as in Ref. [15]. Solving it numerically in one dimension with zero flux boundary conditions and random initial conditions around the equilibrium point (0,0), we reproduce the results reported in Ref. [15], i.e., the system evolves to chaos by the Ruelle-Takens-Newhouse route. The system size is fixed at $L=9.8$ with 50 mesh nodes for both $u$ and $v$. The time step is fixed at $\delta t=0.01$.

Assuming a time series, say $u_{25}$ at the 25*th* node of $u$, measured at $C=-1.3$, as Fig. 4 (a) shows. It is a chaotic trajectory. We apply a segment of $u_{25}$ of length $t=90$ recorded with the interval of $\tau=0.1$ to train the learning machine (1). The control parameters are $c_\beta = 0.05, c_u = 2, c_v = 1, c_b = 150$. The time delay parameter is fixed at $M=100$. Fig. 4(b) shows the bifurcation diagram of the learning machine as a function of $1/\lambda$. For comparison, Fig. 4(c) shows the bifurcation diagram of the BVAM model along the parameter $C$. We see that the two diagrams are highly similar. They show a bifurcation process from period-1 limit cycle to period-2 limit cycle, then to quasiperiodic motion, and chaotic motion. Fig. 4(d) shows several typical trajectories given by the delay coordinate along the two bifurcation diagrams, respectively, confirming that the dynamics do be qualitatively similar. It can be shown from the BVAM model that these solutions represent respectively the period-1 oscillating Turing pattern, two-period oscillating Turing pattern (by period-doubling bifurcation), quasiperiodic oscillating Turing pattern (by Hopf bifurcation), and chaotic oscillation, see Fig. 4(d).

To mimic the situation of a black-box system more realistically we assume that the measured quality is a function of part of system variables, say $\bar{u} = (1/10)\sum_{i=21}^{30} u_i$, we can still obtain the qualitatively same bifurcation diagram, see Fig. 4(e). Therefore, based on only a segment of time series of a measurable quantity of the target system at a fixed parameter set, the learning machine successfully infers the global dynamical properties of the spatiotemporal system in its parameter space. In other words, it correctly predict that the system develops to chaos by the Ruelle-Takens-Newhouse route, as concluded by investing the parameter space of the BVAM model.

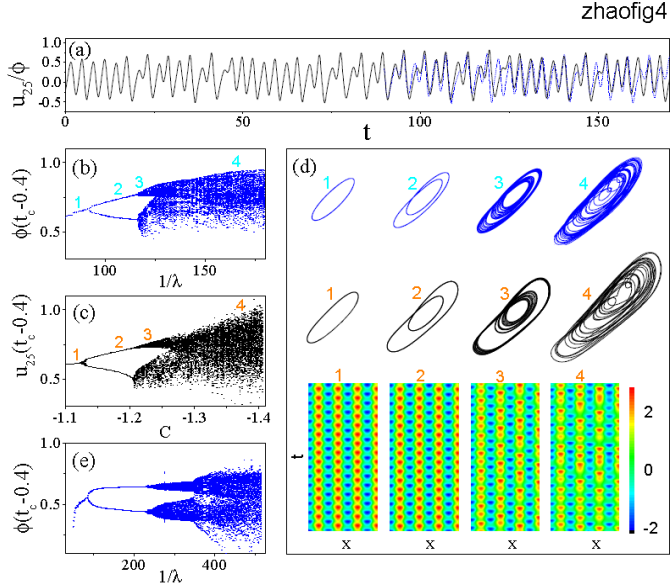

**Fig. 4.** (a) The time series of $u_{25}$ (solid line) made at $C$=-1.3, and the predicted time series of $\phi(t)$ (dashed line). (b) The bifurcation diagram of the learning machine based on the time series of $u_{25}$. (c) The bifurcation diagram of the BVAM model. (d) Typical trajectories along the two bifurcation diagrams. They are represented by $u_{25}(t)$-$u_{25}(t-0.4)$ and $\phi(t)-\phi(t-0.4)$ with scales of $u_{25}(t)/\phi(t) \in (-0.2, 1.1)$, respectively. The plots in the bottom show the spatiotemporal patterns of the BVAM model with respect to these trajectories. The digits tag the positions of these trajectories. (e) The bifurcation diagram of the learning machine based on the time series of $\bar{u}$.

## IV. The mechanism

We first explain why the learning machine can reproduce the dynamics of the target system in the parameter space different from the parameter set that the training data is collected. Note that the Lorenz system can be interpreted as a map, $x(n+1) = x(n) + \tau(-\sigma x + \sigma y)$, $y(n+1) = y(n) + \tau(-xz + Rx - y)$ and $z(n+1) = z(n) + \tau(xy - Bz$, in the Euler form integration. As $\tau \to 0$ these equations tend to the original ones. For the sake of simplicity, we employ a learning machine that covers the Lorenz map as a special case. To this end, we apply (3) as the learning machine and use $f(h) = h^3$ to be the neuron transfer function. Then, it can be written explicitly, for example, for the first equation of motion, it is $\phi_1(n+1) = \phi_1(n) + \tau \sum_{i=1}^{N} u_i (\beta_i(v_{i1}\phi_1(n) + v_{i2}\phi_2(n) + v_{i3}\phi_3(n) - b_i)^3$. The summation part can be expanded into 20 items, with 10 cubic, 6 quadratic, 3 linear ones.

In Fig.5(a) we show the bifurcation diagram of this learning machine when training it with the time series of *x* of the Lorenz system at *B*=0.5725 in the period-3 window. We have applied it to obtain the bifurcation diagram Fig. 2 (a). We see that the bifurcation diagram of the Lorenz system is reproduced up to the period-3 window. In Fig. 5(b) we plot the evolution of the 20 coefficients of the 20 items in the first equation of the learning machine as a function of $1/\lambda$. It is shown that the first equation of the learning machine converges after a short training time to $\phi_1(n+1) = \phi_1(n) + \tau(-\sigma_+\phi_1 + \sigma_-\phi_2)$ approximately, with coefficients of other terms approaching to zero. Though, $\sigma_+ \neq \sigma_-$, the learning machine can be considered as a qualitatively equivalent system of the Lorenz system. It is not exactly the original form but belong to the same class. As a result, after a short stage of training the learning machine collapses to a qualitatively equivalent system of the target system with a parameter set differing from that where the training sample is collected. With the decrease of $\lambda$, $\sigma_+$ and $\sigma_-$ evolve towards $\sigma_+ = \sigma_- = 10$, while coefficients of other terms become smaller and smaller. Consequently, global dynamical behaviors from a simple phase to the present phase are revealed by the learning machine in different training stages.

The learning machine (1) is a time-delayed map. It belongs to a different type of dynamic systems from the Lorenz system. The above approach is not applicable directly. However, the facts revealed by Fig. 1, Fig. 2 and Fig.3 indicate that the training process can also drive this learning machine to a dynamic system that is dynamically equivalent to the target system. Moreover, since it does not converge exactly to the target system, it provides a chance to explore the dynamics over the parameter set that the training data is collected, as Fig.1 (b), Fig. 3(a) and 3(b) show. Therefore, in this sense a learning machine of different type is helpful for inferring the global dynamics of the target system.

Why the learning machine usually first converges to a low-period limit cycle can be easily understood. In the initial stage of the training, the training sample provides qualitative constraints to the learning machine, leading it to collapse to the class of the target system. However, before it accurately approaches the target system that the training data is collected, $x(n+1)$ can be considered as a target of $\phi(n+1)$, and $\phi(n) \equiv x(n)$ is a state vector in the basin of attractor of $x(n+1)$. The cost function (4) just measures the average convergent rate to the attractor. Following the dynamical system theory, if a dynamical system lies in the parameter region with a simple attractor, a period-1 limit cycle for example, the convergent rate should be faster in average comparing to the case lying in parameter region having an attractor with higher complicity, say a period-2 limit cycle. In a chaotic parameter region, though the strange attractor has positive Lyapunov exponents, state vectors in its basin of attractor still converge to the attractor, but has a relatively small convent rate. Therefore, minimizing the cost function causes the learning machine to collapse first to the parameter region having simplest dynamics. As the training progresses further, the finer structure of $x(n+1)$ begins to play a more and more important role, forcing the learning machine evolve towards the present state that the training sample is collected, and reproducing the global dynamics of the target system from simple to complex. The argument for the learning machine (1) is similar.

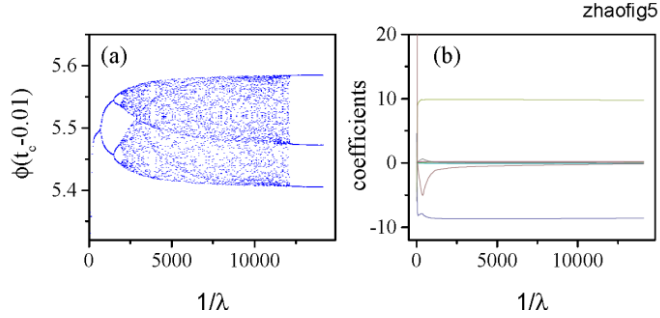

**Figure 5:** (a) The bifurcation diagram of the learning machine with $f(h) = h^3$. (b) The evolution of the 20 coefficients in the first equation of motion of the learning machine.

## V. Conclusion and Discussion

The property the learning machine revealed in this paper provides a model-free framework for inferring the global dynamics of a black-box system. Different from the traditional model-free methods for predicting variable evolutions at the given set of parameters, it reveals the global dynamical properties of the black-box system in its parameter space, usually in the simple-to-complex order. The training sample is a segment of time serious, either chaotic or periodic, collected at a fixed parameter set of the target system. Constrained by the training sample and guided by the cost function, after a short stage of training the learning machine collapses to a qualitatively equivalent system of the black-box system with relatively simple dynamics. In the ensuing evolution, it develops towards the parameter region corresponding to the present state, and thus the dynamics of the target system emerges in the simple-to-complex order.

Based only on a segment of time series obtained at a parameter point, our learning machine correctly infer that the Lorenz system evolves to chaos via the period-doubling route, and the BVAM model develops to chaos via the Ruelle-Takens-Newhouse route. This property of learning machines should be useful for real-world systems that the equations of motion cannot be established accurately. We expect that it provides us a new possibility for predicting the earthquake, epilepsy and atrial fibrillation, etc..

ACKNOWLEDGMENTS

We thank Jiao Wang, Jie Yan, Lamberto Rondoni and Liang Huang for useful discussions. This work is supported by NSFC (Grant No. 11335006).

1. Giraud, O. & Thas, T. Hearing shapes of drums – mathematical and physical aspects of isospectrality, *Reviews of Modern Physics* **82** (3), 2213-2255(2010).
2. John, M. Eigenvalues of the Laplace operator on certain manifolds, *PNAS 51 (4), 542(1964).*
3. Poggio, T., Rifkin, R., Mukherjee, S. & Niyogi, P. General conditions for predictivity in learning theory. *Nature* **428**, 419-422(2004)
4. Tarantola, A. Popper, Bayes and the inverse problem, *Nature Physics* **2**, 492–494(2003)

5. Karianne J. Bergen, Paul A. Johnson, Maarten V. de Hoop, Gregory C. Beroza, Machine learning for data-driven discovery in solid Earth geoscience, *Science* **363**, 1299 (2019)
6. Pilozzi, L., Farrelly, F. A., Marcucci, G. & Conti,C. Machine learning inverse problem for topological photonics. *Communications Physics* **1**, 57(2018)
7. Takens, F. Detecting strange attractors in turbulence. *Dynamical Systems and Turbulence, Warwick 1980*, eds. Rand, D.A. & Young, L.S. (Springer, Berlin).
8. Ma, H.F. *et al.* Randomly distributed embedding making short-term high-dimensional data predictable. *PNAS* **115**, E9994-E1002 (2018).
9. Farmer, J.D. & Sidorowich, J.J. Predicting chaotic time series. *Phys Rev Lett* **59**, 845-848 (1987).
10. Hinton, G.E., Osindero, S. & Teh, Y.W. A fast learning algorithm for deep belief nets. *Neural Comput* **18**, 1527-1554 (2006).
11. Hochreiter, S. & Schmidhuber, J. Long short-term memory. *Neural Comput* **9,** 1735-1780 (1997).
12. Jaeger, H. & Haas, H. Harnessing nonlinearity: Predicting chaotic systems and saving energy in wireless communication. *Science* **304**,78-80(2004).
13. Pathak, J., Hunt, B., Girvan, M., Lu, Z. & Ott, E. Model-free prediction of large spatiotemporally chaotic systems from data: A reservoir computing approach. *Phys Rev Lett* **120**, 024102(2018).
14. Z. Lu, Brian R. Hunt, and E. Ott, Attractor reconstruction by machine learning, Chaos **28**, 061104 (2018).
15. J. L. Arag′on, R. A. Barrio, T. E. Woolley, R. E. Baker, and P. K. Maini, Nonlinear effects on Turing patterns: Time oscillations and chaos, *Phys. Rev. E* **86**, 026201 (2012)
16. Woolley, T. E., Baker, R. E., Maini, P. K. Arag′on, J. L. and Barrio R. A., Analysis of stationary droplets in a generic Turing reaction-diffusion system, *Phys. Rev. E* **82**, 051929 (2010).
17. L. Yang, M. Dolnik, A.M. Zhabotinsky, and I. R. Epstein, Spatial Resonances and Superposition Patterns in a Reaction-Diffusion Model with Interacting Turing Modes, *Phys. Rev. Lett.* **88**, 208303 (2002).
18. V. Petrov, S. Metens, P. Borckmans, G. Dewel, and K. Showalter, Tracking Unstable Turing Patterns through Mixed-Mode Spatiotemporal Chaos, *Phys. Rev. Lett.* **75**, 2895 (1995).
19. J. P. Eckmann, Roads to turbulence in dissipative dynamical systems, *Rev. Mod. Phys.* **53**, 643 (1981).
20. Zhao**,** H. Designing asymmetric neural networks with associative memory. *Phys. Rev. E* **70**, 066137-066143 (2004)
21. H. Zhao, A General Theory for Training Learning Machine, arXiv preprint arXiv:1704.06885 (2017).